\documentstyle[epsfig,float]{aipproc}
\begin{document}

\title{Theoretical Study of the $\gamma \gamma \rightarrow $ Meson-Meson
Reaction}

\author{J.A. Oller and E. Oset}
\address{ Departamento de F\'{\i}sica Te\'orica and IFIC\\
Centro Mixto Universidad de Valencia-CSIC\\
46100 Burjassot (Valencia), Spain}

\maketitle

\begin{abstract}
We present a unified theoretical picture which studies simultaneously
the $\gamma \gamma \rightarrow \pi^+ \pi^-$, $\pi^0 \pi^0$,
$K^+ K^-$, $K^0 \bar{K}^0$, $\pi^0 \eta$ reactions up to about $\sqrt{s} = 1.4 \,\,
 GeV$ reproducing the experimental cross sections . The
present work implements in an accurate way the final state interaction
of the meson-meson system, which is shown to be 
essential in order to reproduce the data, particularly the $L = 0$ channel.
This latter channel 
is treated here following a recent theoretical work in
which the meson-meson
 amplitudes are well reproduced and the $f_0, a_0, \sigma$
resonances show up clearly as poles of the $t$ matrix. The present work,
as done in earlier ones,
also incorporates elements of chiral symmetry and exchange of vector and
axial resonances in the crossed channels, as well as a direct coupling to the
$f_2 (1270)$ and $a_2 (1370)$ resonances. We also evaluate the decay 
width of the $f_0 (980)$ and $a_0 (980)$ resonances into the $\gamma 
\gamma$ channel.  
\end{abstract}

\section{Introduction}

The $\gamma \gamma \rightarrow$ meson-meson reaction 
provides interesting information concerning the structure of hadrons,
their spectroscopy and the meson-meson interaction, given the sensitivity 
of the reaction to the hadronic final state interaction (FSI)
\cite{10}. 

 The chiral perturbation approach \cite{18,Bij} is valid only  
for low energies. However, chiral
symmetry is one of the important ingredients when dealing with the
meson-meson interaction. In addition to the $f_0 (980), a_0 (980)$ 
and $ \sigma$ resonances, which come up naturally in the approach 
of ref. \cite{15}, which we use here, we introduce phenomenologically 
the $f_2 (1270) $ and 
$a_2 (1320)$ resonances.Another relevant 
aspect of this reaction, which has been reported previously,
is the role of the vector and axial resonance exchange in the $t$, $u$
channels. 

\section{Final state interaction corrections}

The vertex contributions come from the Born term, only for charged particles, 
and the exchange in the t,u channels of axial and vector resonances, \cite{18}.
We now take into account their FSI corrections.

{\bf$S$-wave.}

The one loop contribution generated from the Born terms with 
intermediate charged mesons can be directly taken from $\chi P T$ 
calculations of the $\gamma \gamma \rightarrow \pi^0 \pi^0$ amplitude 
to one loop. 
The important point is that the strong amplitude connecting 
the charge particles with the $\pi^0\pi^0$ factorizes. Our contribution 
beyond this point is to include all meson loops generated by
the coupled channel Lippmann Schwinger equations of ref. \cite{15}, in
which we also saw that the on shell meson-meson amplitude factorizes
outside the loop integrals. Hence, the immediate consequence of
introducing these loops is to substitute the on shell $\pi \pi$ amplitude
of order $O (p^2)$ by our on shell meson-meson amplitude
evaluated in ref. \cite{15}. A similar procedure to account 
for FSI in the terms generated by vector and axial resonance exchange 
can be applied iterating the potential in the meson-meson amplitude.
\begin{equation}
\begin{array}{l}
t_{R, M_3 M_4} = \sum_{M_1 M_2} \tilde{t}_{R, M_1 M_2} t_{M_1 M_2, M_3 M_4} \\
\tilde{t}_{R, M_1 M_2} =i \int \frac{d^4 q}{(2 \pi)^4} t'_R,_{M_1 M_2}
 \, \frac{1}{q^2 - m_1^2 + i \epsilon}
 \, \, \frac{1}{(P - q)^2 - m_2^2 + i \epsilon}
\end{array}
\end{equation}

 In eq. (3) $P = (\sqrt{s},\vec{0})$ is
the total fourmomentum of the $\gamma \gamma$ system and $m_1, m_2$
the masses of the intermediate $M_1, M_2$ mesons. In addition, and in 
analogy to the work of ref. \cite{15}, the integral over $|\vec{q}|$ 
in eq. (3) is 
cut at $q_{max} = 0.9 \, GeV$.One can justify the accuracy of factorizing 
the strong amplitude for the loops with crossed exchange of resonances \cite{
Oller}.

{\bf D-wave contribution}

For the $(2, 2)$ component we take the results of ref. \cite{26}, obtained
using dispersion relations
\vspace{-0.5cm}
\begin{equation}
t_{BC}^{(2,2)} = \biggl[ \frac{2}{3} \, \chi_{22}^{T=0} \, e^{i \delta_{20}} +
 \frac{1}{3} \, \chi_{22}^{T=2} \, e^{i \delta_{22}} \biggr]\, t_B^{(2, 2)}
\end{equation}
For the $\gamma \gamma \rightarrow K^+ K^-$ reaction the non resonant 
D-wave contribution is not needed because we are close to $K\bar{K}$ 
threshold and furthermore the functions $\chi_{ij}$ 
are nearly zero close to the mass of the $f_2$ and $a_2$ 
resonances.

The {\bf resonance contribution} in the $D-$wave coming form the 
$f_2 (1270)$ and $a_2 (1320)$ resonances is parametrised in the standard 
way of a Breit-Wigner as done in ref. \cite{5}. The parameters of these 
resonances are completely compatible with the ones coming from the Particle 
Data Group \cite{36}
\vspace{-.2cm}
\section{Results}
\vspace{-.2cm}

{\bf Total and differential cross sections }

\begin{figure}[H]

\centerline{
\protect
\hbox{
\psfig{file=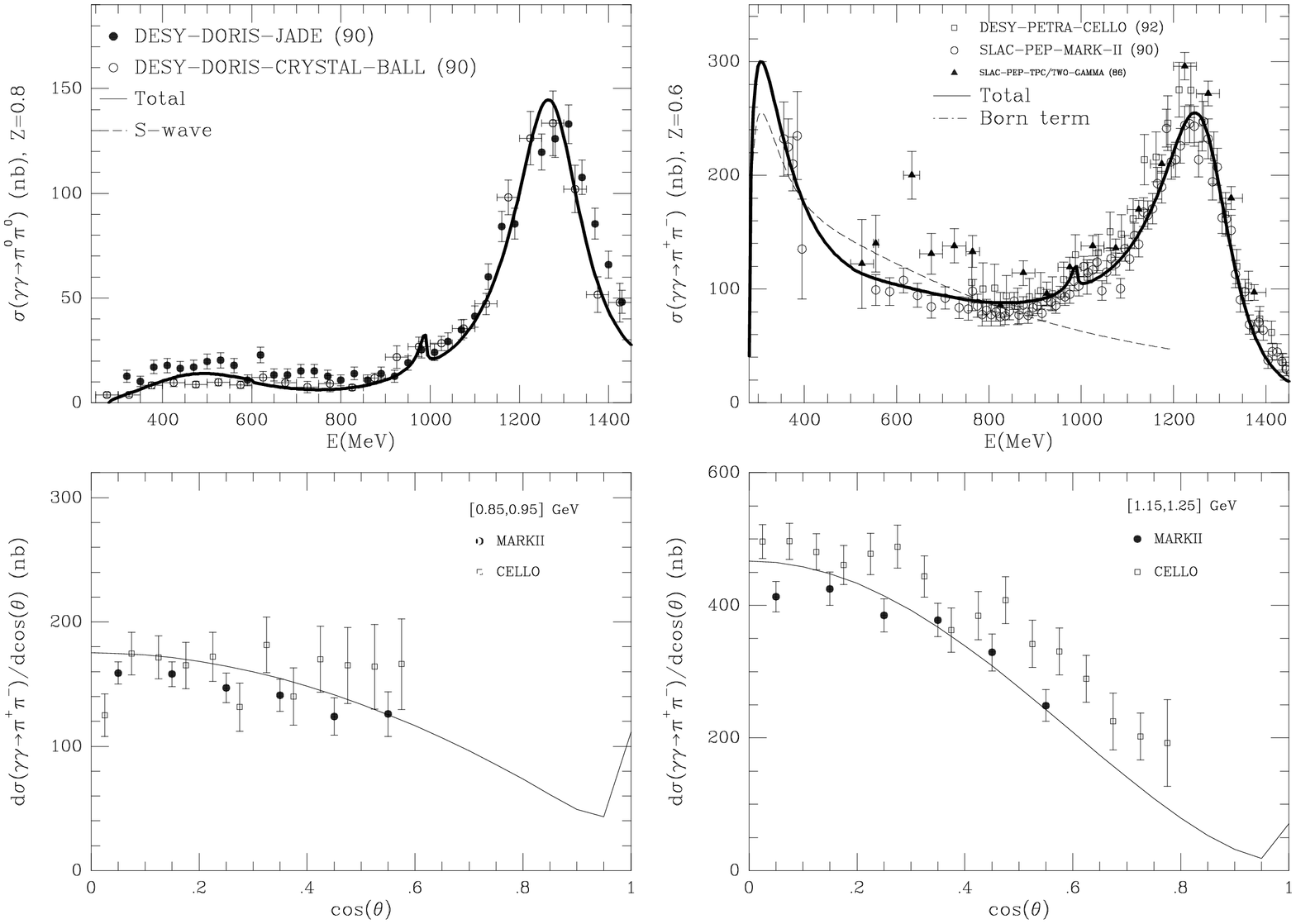,width=0.7\textwidth,angle=-90}}}
\end{figure}

\begin{figure}[H]
\centerline{
\protect
\hbox{
\psfig{file=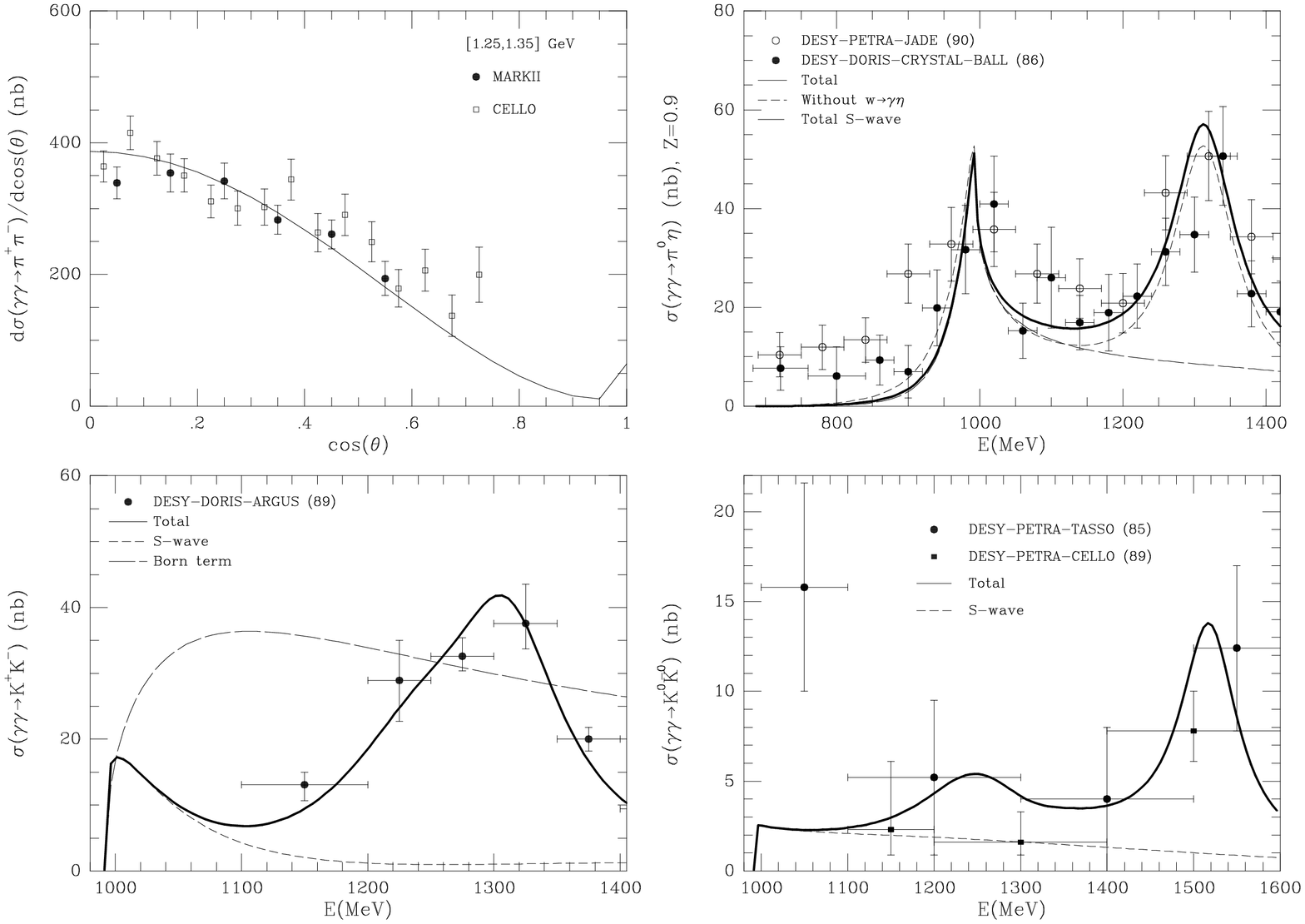,width=0.7\textwidth,angle=-90}}}
\end{figure}

{\bf Partial decay width to two photons of the $f_0 (980)$
and $a_0 (980)$.}

We follow the same scheme than in \cite{15} in order to use directly our 
calculated strong and 
photo-production amplitudes. From our amplitudes in isospin $T = 1$ and  for the
isospin $T = 0$ part, taking the terms which involve the strong
$M \bar{M} \rightarrow M \bar{M}$ amplitude, we isolate the part of the
$\gamma \gamma \rightarrow M \bar{M}$ which proceeds via the resonances $a_0$
and $f_0$ respectively. In the vicinity of the resonance the amplitude 
proceeds
as $M \bar{M} \rightarrow R \rightarrow M \bar{M}$. Then we eliminate
the $R \rightarrow M \bar{M}$ part of the amplitude plus the $R$ 
propagator and remove the proper
isospin Clebsch Gordan coefficients for the final states (1 for $\pi^0 \eta$
and $- 1/ \sqrt{2}$ for $K^+ K^-$) and then we get the coupling of the 
resonances to the $\gamma \gamma$ channel.
\begin{equation}
\begin{array}{lll}
\Gamma_{a_0}^{\gamma \gamma} = 0.78 \; KeV &
\Gamma_{a_0}^{\gamma \gamma} 
\frac{\Gamma_{a_0}^{\eta \pi}}{\Gamma_{a_0}^{tot}}
= 0.49 \; \, KeV &
\Gamma^{\gamma \gamma}_{f_0} = 0.20 \;  \, KeV
\end{array}
\end{equation}
 
\section{Conclusions}

1) The resonance $f_0 (980)$ shows up weakly in 
$\gamma \gamma \rightarrow \pi^0 \pi^0$ and barely in 
$\gamma \gamma \rightarrow \pi^+ \pi^-$.

2) In order to explain the angular distributions of the
$\gamma \gamma \rightarrow \pi^+ \pi^-$  reaction we did not need the
hypothetical $f_0 (1100)$ broad resonance suggested in other works \cite{26}.
 This also solves the puzzle of why it did not show up in the 
$\gamma \gamma \rightarrow \pi^0 \pi^0$ channel. Furthermore, such resonance
does not appear in the theoretical work of ref. \cite{15}, while the 
$f_0 (980)$ showed up clearly as a pole of the $t$ matrix in $T = 0$.

3) The resonance $a_0$ shows up clearly in the 
$\gamma \gamma \rightarrow \pi^0 \eta$ channel and we reproduce 
the experimental results without the need of an extra background 
from a hypothetical $a_0(1100-1300)$ resonance suggested in ref. 
\cite{10}.

4) We have found an explanation to the needed reduction of the Born term
in the $\gamma \gamma \rightarrow  K^+ K^-$  reaction in terms of
final state interaction of the $K^+ K^-$ system.

\end{document}